\documentclass[cameraready]{Interspeech}
\usepackage{amsmath}
\usepackage{colortbl}
\usepackage{multirow}
\usepackage{bm} 
\usepackage{cite}

\newcommand\blfootnote[1]{%
  \begingroup
  \renewcommand\thefootnote{}\footnote{#1}%
  \addtocounter{footnote}{-1}%
  \endgroup
}

\definecolor{pastelviolet}{RGB}{234, 229, 246}

\title{VIB-AVSR: Variational Information Bottleneck for Noise-Robust LLM-Based Audio-Visual Speech Recognition}

\author[affiliation={\heartsuit,*}]{Piyush}{Arora}
\author[affiliation={\heartsuit,*}]{Navlika}{Singh}
\author[affiliation={\heartsuit}]{Umberto}{Cappellazzo}
\author[affiliation={\heartsuit,\spadesuit}]{Stavros}{Petridis}
\author[affiliation={\heartsuit,\spadesuit}]{Maja}{Pantic}

\address{
    $^\heartsuit$ Imperial College London, UK \\
    $^\spadesuit$ NatWest AI Research, UK
}

\email{pa524@ic.ac.uk, ns1324@ic.ac.uk}


\usepackage{comment}

\begin{document}

\maketitle




\begin{abstract}
Audio-Visual Speech Recognition takes two input modalities, acoustic and visual streams, where visual information from lip movements aids recognition when audio is noisy. Recently, LLM-based AVSR models have emerged as a promising paradigm by connecting pre-trained audio-visual encoders to an LLM, achieving strong results in clean conditions. However, these models are predominantly optimized for clean acoustic conditions, with limited attention to making the LLM backbone robust to noise. No explicit mechanism is employed to produce stable representations under corrupted audio, leading to performance degradation in noisy environments. To address this, we propose VIB-AVSR, which integrates Variational Information Bottleneck layers at targeted positions within the LLM backbone to regularize representations. VIB-AVSR reduces degradation under noisy conditions across multiple SNR levels and noise types, without requiring architectural modifications or additional training data. Our code is available at \href{https://github.com/PiyushArora1010/VIB-AVSR}{https://github.com/PiyushArora1010/VIB-AVSR}.
\blfootnote{* denotes equal contribution.}
\end{abstract}

\section{Introduction}
Audio-Visual Speech Recognition (AVSR) improves transcription accuracy by jointly processing acoustic and visual streams, where lip movements provide a complement to the audio signal. The field has progressed through a series of end-to-end deep learning approaches based on Conformer and Transformer backbones~\cite{petridis2018audio, ma2021end, serdyuk2022transformer,burchi2023audio}, self-supervised pre-training methods such as AV-HuBERT~\cite{shi2022robust, shilearning} and large-scale automatic labeling pipelines such as Auto-AVSR~\cite{ma2023auto}, which have driven consistent improvements on standard benchmarks. More recently, approaches leveraging large-scale pre-trained ASR models, such as Whisper-Flamingo~\cite{rouditchenko2024whisper, li2026noise, rouditchenko2025mwhisper}, have demonstrated strong noise robustness by injecting visual features into a pre-trained speech encoder and the decoder. Building on these advances, LLM-based AVSR models~\cite{cappellazzo2025large, yeo2024visual, cappellazzo2025scaling, cappellazzo2025mome, yeo2025mms, cappellazzo2025omni, cappellazzo2025adaptive, anand2026mitigating, cappellazzo2026dr} have emerged as a promising paradigm, connecting pre-trained audio and video encoders to a Large Language Model (LLM) via lightweight adapters, achieving state-of-the-art performance.

Despite these advances, noise robustness remains an largely overlooked challenge in the LLM-based AVSR paradigm. LLM-based AVSR models are predominantly optimized for clean acoustic conditions, and prior work has noted a non-trivial performance gap when these models are evaluated under noisy conditions~\cite{cappellazzo2025large, cappellazzo2025scaling, yeo2025mms}. This stands in contrast to traditional encoder-decoder AVSR models, which are trained end-to-end from scratch and can therefore develop noise-robust representations throughout the entire architecture~\cite{shi2022robust, hong2023watch, kim2025mohave, kim2025multi}. In LLM-based AVSR, the backbone is a pre-trained language model whose parameters have been optimized purely on text, it has never been exposed to noisy audio-visual speech, and only a small number of its parameters are updated during fine-tuning via LoRA~\cite{hu2022lora}. As a result, no explicit mechanism exists within the LLM backbone to produce stable representations when the input audio is corrupted, and the burden of noise robustness falls entirely on the encoders, leaving the LLM itself ill-equipped to handle the domain shift induced by acoustic noise. We argue that addressing this requires a principled approach to regularizing the LLM's internal representations directly, rather than relying solely on data augmentation or encoder-only modifications.

Motivated by this, we propose VIB-AVSR, a novel framework that inserts Variational Information Bottleneck (VIB)~\cite{alemi2017deep} layers at targeted positions within the LLM backbone. The VIB objective encourages the model to learn representations that are maximally informative about the transcription target while discarding noise-induced variance, directly addressing the vulnerability at its source. Crucially, VIB-AVSR requires no major architectural changes, no additional training data, and introduces only negligible computational overhead, making it a lightweight and practical addition to existing LLM-based AVSR pipelines. In summary, our contributions are as follows:

\begin{itemize}
    \item We propose VIB-AVSR, a lightweight method that integrates VIB layers into the LLM backbone of an AVSR model to improve noise robustness at the representation level, without requiring architectural redesign or additional training data.
    \item We demonstrate that VIB-AVSR improves noise robustness under both noisy and clean training paradigms, showing that variational compression promotes generalisation to noisy conditions.
    \item We conduct extensive ablation studies on VIB layer placement, regularisation strength $\beta$, and interpolation coefficient $\alpha$, providing empirical evidence that supports each design choice and identifying the most effective configuration.
    \item VIB-AVSR achieves WER reductions over Llama-AVSR across multiple noise types and SNR levels, with gains that widen under extreme noise conditions, while preserving recognition performance on clean speech.
\end{itemize}

\begin{figure}[t]
    \centering
        \includegraphics[width=0.98\columnwidth, trim=0 0 80 0, clip]{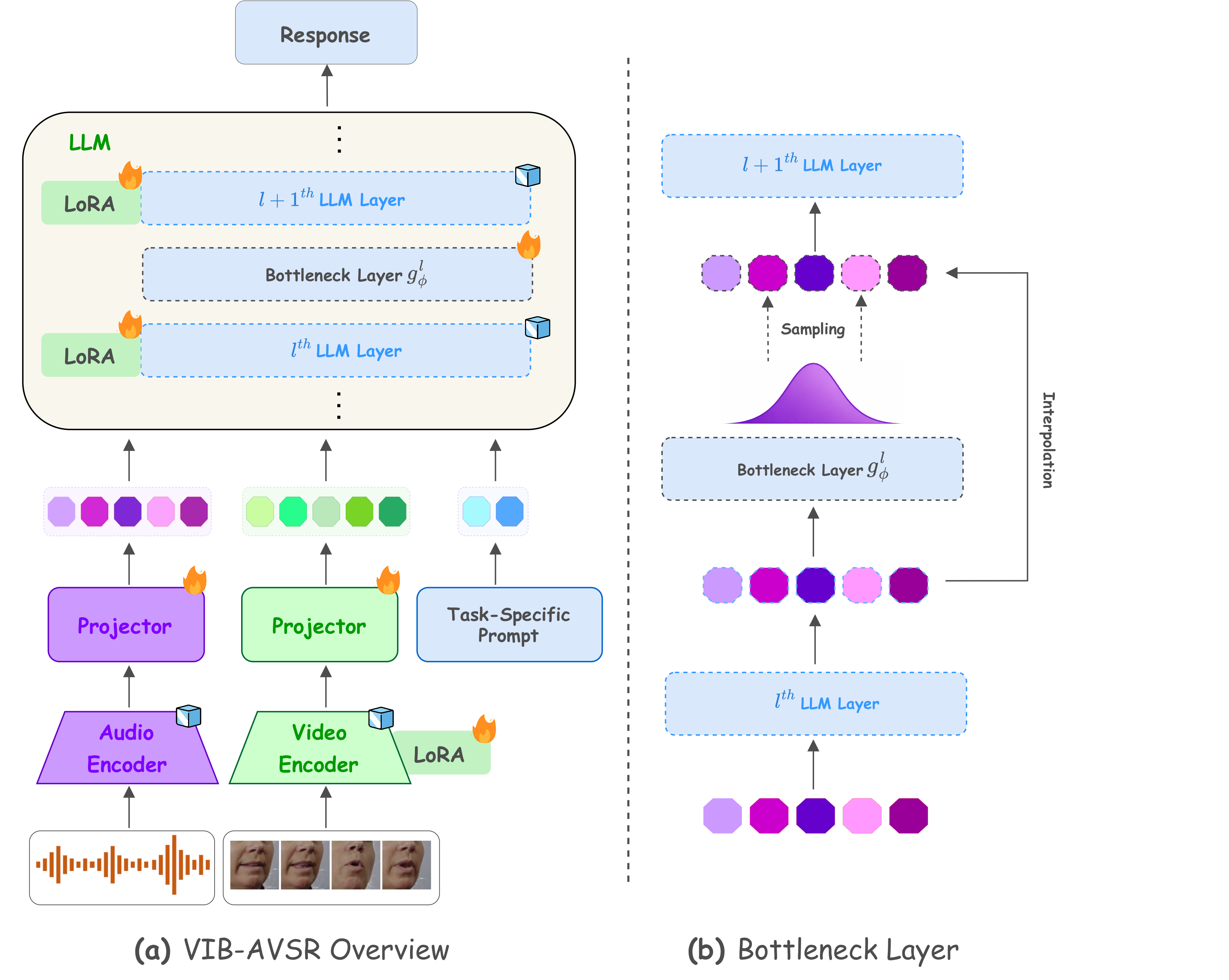}
    \caption{Architecture overview of VIB-AVSR. (a) Audio and video inputs are encoded and projected into a shared representation that conditions an LLM augmented with LoRA adapters and variational information bottleneck (VIB) module. (b) The bottleneck layer compresses intermediate representations, performs latent sampling, and interpolates features before passing them to the next LLM layer.}
    \label{fig:method_overview}
    \vspace{-0.5cm}
\end{figure}
\section{Background}
\textbf{Llama-AVSR:} Cappellazzo et al.~\cite{cappellazzo2025large} proposed Llama-AVSR, which is a Multimodal Large Language Model (MLLM) for audio, visual, and audio-visual speech recognition. It consists of three components: modality-specific pre-trained encoders, lightweight linear projectors, and a pre-trained LLM backbone. An audio encoder and a video encoder independently process their respective input streams, producing feature sequences that are downsampled and projected into the LLM's embedding space via modality-specific linear projectors. The resulting audio tokens ${H}_a$, video tokens ${H}_v$, and text tokens ${H}_t$ are concatenated and processed by the LLM, which generates a transcription of the audio auto-regressively.

Let $X = (X^a, X^v, X^t)$ denote the multimodal input and $Y$ the target transcription, the model is trained with a standard autoregressive objective:
\begin{equation*}
\arg\min_\theta \, \mathbb{E}_{X,Y}
\left[
\sum_{m=1}^{M}
-\log f_\theta(Y_m \mid X^a, X^v, X^t, Y_{<m})
\right]
\end{equation*}
where $\theta$ denotes the trainable parameters, i.e. projection layers and LoRA modules, $M$ is the transcription length, $Y_{<m} = \{Y_1, \dots, Y_{m-1}\}$ denotes generated transcript up to token $m-1$, and $f_\theta(\cdot)$ is the model's predicted probability distribution. Although this formulation achieves strong performance under clean acoustic conditions, the audio representations are sensitive to acoustic noise, as the model does not explicitly regularize against noise-induced variation in the audio hidden states.

\noindent\textbf{Information Bottleneck (IB) principle:} Tishby et al.~\cite{tishby2000information} introduces a framework for learning representations that retain task-relevant information while discarding redundant variation. For an input variable $X$, it's learned representation $Z$ parameterized by $\phi$, and a target $Y$, the IB objective is defined as:
\begin{equation}
\max_{\phi} \ I(Z; Y) - \beta I(Z; X)
\end{equation}
where $I(\cdot\,;\cdot)$ represents mutual information and $\beta \ge 0$ controls the compression-prediction trade-off. Maximising $I(Z;Y)$ preserves information predictive of the output, while minimising $I(Z;X)$ discards redundancy in input, such as acoustic noise. In the AVSR setting, noise corrupts the audio input $X^a$, causing the audio hidden states ${H}_a$ to encode noise-specific features that degrade transcription. We apply the IB objective selectively to ${H}_a$ to compress noise-related variation while retaining speech-discriminative content. The video and text hidden states ${H}_v$ and ${H}_t$ are not bottlenecked, as they are not directly corrupted by acoustic noise and should be preserved to provide complementary lip-movement and linguistic context to the LLM.

\section{VIB-AVSR}
Now, a direct optimisation of the IB objective is intractable due to the difficulty of estimating mutual information in high-dimensional spaces. Following the variational formulation~\cite{alemi2017deep, ohvisual}, we derive a tractable lower bound tailored to the audio hidden states in Llama-AVSR.



The audio encoder and the respective projector presents the initial audio representations ${H}^0_a$, which are passed into the LLM. The LLM backbone consists of $L$ transformer layers, and the audio hidden states at the output of layer $l$ are denoted ${H}^l_a \in \mathbb{R}^{T \times d}$, where $T$ is the number of audio tokens and $d$ is the hidden dimension. Without loss of generality, we derive the bottleneck formulation for a single layer $l$, where the variational information bottleneck (VIB) module $g_\phi^l$ takes ${H}^l_a$ as input and produces a compressed representation $Z^l_a$.

For the compression term $I(Z^l_a;{H}^l_a)$, we introduce a factorizable variational prior $r(Z^l_a)$ and apply the non-negativity of KL divergence to obtain:
\begin{equation}
I(Z^l_a; {H}^l_a) \le
\mathbb{E}_{{H}_a^l} \big[
D_{\mathrm{KL}}(
p(Z^l_a \mid {H}^l_a) \, || \, r(Z^l_a)
)
\big]
\end{equation}
For the predictive term $I(Z^l_a; Y)$, we substitute the true posterior $p(Y|Z^l_a)$ with a variational approximation $q_\phi(Y |Z^l_a, X^v, X^t)$, parameterised by the LLM decoder. Omitting the constant $H(Y)$ with respect to model parameters, this gives the lower bound:
\begin{equation}
I(Z^l_a;\, Y) \ge
\mathbb{E}_{X,Y}
\left[
\mathbb{E}_{Z^l_a \mid {H}^l_a}
\left[
\log q_\phi(Y \mid Z^l_a, X^v, X^t)
\right]
\right]
\end{equation}
Combining both bounds yields the variational IB objective at layer $l$, which is maximised with respect to the model parameters:
\begin{align}
\label{eq:vib_objective}
\mathcal{L}^l_{\mathrm{VIB}} &=
\mathbb{E}_{X,Y} \Big[
\mathbb{E}_{Z^l_a \mid {H}^l_a} \Big[
\log q_\phi(Y \mid Z^l_a, X^v, X^t)
\Big]
\Big] \nonumber \\
&\quad - \beta \,
\mathbb{E}_{{H}^l_a} \Big[
D_{\mathrm{KL}}(
p(Z^l_a \mid {H}^l_a) \, || \, r(Z^l_a)
)
\Big]
\end{align}
The first term is the autoregressive transcription likelihood and the second term regularises the audio hidden states by penalising deviation from the prior. The hyperparameter $\beta$ controls the compression-prediction trade-off.

\begin{table*}[t]
\centering
\caption{WER (\%) of Llama-AVSR and VIB-AVSR under babble and speech noise at five SNR levels, evaluated under both noisy and clean training paradigms. Avg~($N$$>$$S$) denotes the average WER over conditions where noise dominates the signal ($-10$, $-5$, $-2$ dB). $\infty$ denotes noise-free evaluation. Bold indicates the best result per configuration.}
\label{tab:wer_comparison}
\scriptsize
\setlength{\tabcolsep}{4pt}
\renewcommand{\arraystretch}{1.1}

\resizebox{\textwidth}{!}{
\begin{tabular}{cc *{7}{c} *{7}{c} c}
\toprule

\multirow{3}{*}{Training} & \multirow{3}{*}{Method}
& \multicolumn{7}{c}{Babble} & \multicolumn{7}{c}{Speech} &  \\

\cmidrule(lr){3-9} \cmidrule(lr){10-16}

& 
& -10 & -5 & -2 & 0 & 5 & Avg & Avg ($N$$>$$S$)
& -10 & -5 & -2 & 0 & 5 & Avg & Avg ($N$$>$$S$)
& \multirow{1}{*}{$\infty$}\\

\midrule

\multirow{2}{*}{Noisy}
& Llama-AVSR
& 34.87 & 27.55 & 18.09 & 8.00 & 5.74 & 18.85 & 26.84
& 27.66 & 14.94 & 9.14 & 5.21 & \textbf{3.84} & 12.16 & 17.24
& 2.72 \\

& \underline{VIB-AVSR}
& \textbf{32.52} & \textbf{25.63} & \textbf{15.63} & \textbf{7.82} & \textbf{5.35} & \textbf{17.39} & \textbf{24.59}
& \textbf{27.52} & \textbf{13.23} & \textbf{8.90} & \textbf{5.08} & 3.98 & \textbf{11.74} & \textbf{16.55}
& \textbf{2.38} \\

\midrule

\multirow{2}{*}{Clean}
& Llama-AVSR
& 47.03 & 34.32 & 20.20 & 8.61 & 5.19 & 23.07 & 33.85
& 42.13 & 18.44 & 10.78 & \textbf{5.30} & \textbf{4.32} & 16.20 & 23.78
& \textbf{2.34} \\

& \underline{VIB-AVSR}
& \textbf{40.97} & \textbf{31.50} & \textbf{19.45} & \textbf{8.55} & \textbf{5.00} & \textbf{21.09} & \textbf{30.64}
& \textbf{37.44} & \textbf{17.37} & \textbf{10.69} & 5.36 & 4.39 & \textbf{15.05} & \textbf{21.83}
& 2.42 \\

\bottomrule

\end{tabular}
}
\vspace{-0.5cm}
\end{table*}

\noindent\textbf{Practical Implementation:} Using a Monte Carlo approximation 
of the expectations over data and negating Eq.~(\ref{eq:vib_objective}) 
to form a minimisation objective, the empirical estimate over a 
mini-batch of $N$ samples is derived as:
\begin{align}
\mathcal{L}_\beta 
&= \frac{1}{N} \sum_{i=1}^{N}
\Bigg[
\mathbb{E}_{Z^{l,i}_a \mid H^{l,i}_a}
\left[
-\log q_\phi\left(Y^i \mid Z^{l,i}_a, X^{v,i}, X^{t,i}\right)
\right] \nonumber \\
&\hspace{5em}
+ \beta \,
D_{\mathrm{KL}}\left(
p\left(Z^{l,i}_a \mid H^{l,i}_a\right)
\,||\,
r\left(Z^{l}_a\right)
\right)
\Bigg]
\label{eq:vibavsr}
\end{align}
where $Z^{l}_a$ denotes the bottlenecked audio representation at 
layer $l$. The first term maximises the likelihood of the target transcription under the compressed representation, while the KL term 
regularises the posterior $p(Z^{l}_a \mid H^{l}_a)$ towards the 
prior $r(Z^{l}_a)$. Both distributions are modelled as diagonal 
Gaussians~\cite{kingma2022autoencodingvariationalbayes}.

\noindent\textbf{Prior distribution.} 
The prior is a learnable diagonal Gaussian,
\begin{equation}
r(Z^{l}_a) = \mathcal{N}\left( Z^{l}_a;\, {\mu}^{l}_r,\, 
({\sigma}^{l}_r)^2 \cdot I \right)
\end{equation}
where ${\mu}^{l}_r \in \mathbb{R}^d$ and ${\sigma}^{l}_r \in 
\mathbb{R}^d_+$ are per-layer parameters shared across all samples, 
admitting a closed-form KL divergence.

\noindent\textbf{Posterior distribution.} 
We parameterise the posterior using a position-wise two-layer MLP
$g_{\phi} : \mathbb{R}^d \rightarrow \mathbb{R}^{2d}$,
applied independently to each audio token embedding, analogously to 
the feed-forward sublayer of a Transformer~\cite{vaswani2017attention}. 
Given ${H}^{l}_a$ at layer $l$:
%
$
p(Z^{l}_a \mid {H}^{l}_a)
:= \mathcal{N}(Z^{l}_a;\, {\mu}^{l},\, ({\sigma}^{l})^2 \cdot I)$
%
where $[ {\mu}^{l},\, ({\sigma}^{l})^2 ] = g_{\phi}({H}^{l}_a)$,
with mean and variance partitioned along the output dimension.
The MLP is applied exclusively to audio token positions, leaving 
visual and text representations unchanged. Samples from the posterior 
are obtained via the reparameterisation 
trick~\cite{kingma2022autoencodingvariationalbayes}:
\begin{equation}
\tilde{Z}^{l}_a =
{\mu}^{l}
+ {\sigma}^{l} \odot {\epsilon},
\quad
{\epsilon} \sim \mathcal{N}(\mathbf{0}, I)
\end{equation}
During inference, we follow the standard practice~\cite{kingma2022autoencodingvariationalbayes} and instead of reparameterization trick, only ${\mu}^{l}$ is used. To balance compression with the retention of speech-discriminative content, the bottleneck output 
is interpolated with the pre-bottleneck 
representation~\cite{ohvisual}:
%
$
\hat{Z}^{l}_a = \alpha  {H}^{l}_a + (1-\alpha) \tilde{Z}^{l}_a
$
%
where, $\alpha$ is set to 0.5 for our experiments. This interpolated representation $\hat{Z}^{l}_a$ 
replaces ${H}^{l}_a$ and is propagated to the subsequent LLM layers. In practice, 
we experiment with inserting bottleneck modules after multiple layers~\cite{ohvisual} in the LLM, with each layer maintaining its own independent prior. Early experiments with $\alpha = 0$ and with scheduling $\alpha$ from $0$ to $0.5$ following~\cite{ohvisual} both resulted in higher WERs, which we speculate is due to the sampled representation being too lossy for the LLM to generate.

\section{Experiments and Discussion}

\subsection{Implementation Details}

\noindent\textbf{Dataset.} We train and evaluate on the LRS2~\cite{son2017lip} dataset, consisting of BBC program clips with transcribed English speech.

\noindent\textbf{Model Architecture.} Following Llama-AVSR~\cite{cappellazzo2025large}, we adopt a multimodal LLM framework consisting of three components: a pre-trained audio encoder, a pre-trained video encoder, and an LLM backbone. We use Whisper-medium~\cite{radford2023robust} as the audio encoder and AV-HuBERT~\cite{shilearning} as the video encoder. Modality-specific linear projectors are trained to map audio and video features to the LLM embedding space. Llama-3.2-1B~\cite{grattafiori2024llama} is incorporated as the LLM backbone. The compression rates for audio and video tokens is set to 3 each. The same configuration is used to train Llama-AVSR for fair comparison.

\noindent\textbf{Training Details.} The LLM and video encoder are fine-tuned using LoRA~\cite{hu2022lora}, with rank 64 for the LLM and rank 16 for the video encoder. The audio encoder is kept frozen throughout the training. VIB modules are inserted after selected LLM layers and trained jointly with the LoRA modules. The insertion points and the value of $\beta$ are chosen based on ablation studies over layer placement, bottleneck count, and regularization strength (Section~\ref{section:ablations}), which identify layers 4 and 8 with $\beta = \frac{0.1}{H}$ as the most effective configuration, where $H$ denotes the LLM's hidden dimension size. Each VIB module $g_\phi^l$ is implemented as a two-layer MLP: a linear layer followed by GeLU activation and a second linear layer with $2H$ output neurons, from which the mean and log-variance of the variational posterior are obtained.

\noindent\textbf{Training Paradigms.} For the experiments, we mainly evaluate two training paradigms. In the Clean Paradigm, no noise is added to the audio during training. In the Noisy Paradigm, a random noise is added to the training audio, with the SNR level sampled uniformly at random per instance. Evaluating VIB-AVSR under both paradigms allows us to assess whether the variational bottleneck improves robustness when noise is seen during training, and whether its regularisation effect generalises to noisy conditions even when trained on clean conditions.

\subsection{Main Results}

Table~\ref{tab:wer_comparison} reports WER for Llama-AVSR and VIB-AVSR across babble and speech noise, sampled from the MUSAN~\cite{musan} dataset at five SNR levels, under both noisy and clean training paradigms. We additionally report extreme noise average Avg~($N$$>$$S$), the average WER over SNR levels where noise substantially dominates the signal ($-10$, $-5$, $-2$ dB), as a measure of robustness under extreme acoustic degradation.


\noindent\textbf{Noisy Training.} Under noisy training, VIB-AVSR almost always reduces WER relative to Llama-AVSR, with the largest gains observed at lower SNRs, particularly under babble noise. This trend is further reflected in the Avg~($N$$>$$S$) metric. These results suggest that the compression objective becomes increasingly beneficial as acoustic interference dominates the speech signal. While improvements under speech noise are smaller than those observed for babble noise, VIB-AVSR still reduces WER across all SNR conditions except 5 dB. Notably, VIB-AVSR also achieves lower WER than Llama-AVSR under noise-free evaluation ($\infty$). This indicates that the bottleneck acts as an effective regularizer, encouraging representations that retain task-relevant information while suppressing nuisance variability beyond the noisy conditions.


\noindent\textbf{Clean Training.} When trained without noise augmentation, VIB-AVSR still yields substantial WER reductions. Under babble noise, improvements are across all SNR levels, while under speech noise VIB-AVSR provides consistent gains in the low-SNR regime. Notably, the bottleneck is never exposed to noisy samples during training, yet the learned representations generalise more effectively to unseen noisy conditions. This suggests that variational compression promotes robustness through a mechanism distinct from data augmentation. The effect is especially evident in the Avg~($N$$>$$S$) metric. As expected, models trained on clean data degrade more sharply under extreme noise than their noisy-trained counterparts, reflecting the impact of train-test domain mismatch. However, VIB-AVSR consistently narrows this gap relative to Llama-AVSR, particularly in the lowest-SNR conditions where the baseline is most vulnerable, indicating that the bottleneck partially compensates for the absence of noise augmentation. Finally, performance on clean speech ($\infty$) remains comparable, demonstrating that the robustness gains do not come at the cost of clean-conditions.

\subsection{Ablation Studies}
\label{section:ablations}

\begin{table}[t]
\centering
\caption{WER (\%) under babble for VIB insertions. The best results are in bold and the second-best are underlined.}
\label{tab:vib_layers}
\scriptsize
\setlength{\tabcolsep}{4pt}
\renewcommand{\arraystretch}{1.2}

\resizebox{0.83\columnwidth}{!}{
\begin{tabular}{c *{6}{cc}}
\toprule
& \multicolumn{6}{c}{SNR (dB)} \\
\cmidrule(lr){2-7}
Layer(s) & $-10$ & $-5$ & $-2$ & $0$ & $5$ & $\infty$ & Avg \\
\midrule
\rowcolor{gray!20}
\multicolumn{8}{l}{\textbf{Single}} \\
-1 & \underline{33.25} & \underline{25.27} & 16.29 & \underline{7.33} & 5.59 & \underline{2.37} & 15.03 \\
2      & 34.76 & 25.59 & 16.95 & \textbf{7.12} & 5.20 & 2.43 & 15.32 \\
4      & 34.73 & {25.50} & 16.95 & 8.17 & \underline{5.18} & 2.61 & 15.52 \\
6      & 34.59 & 26.32 & 16.85 & 7.48 & 5.36 & 2.76 & 15.56 \\
8      & 34.53 & 26.44 & 17.67 & 7.55 & \textbf{5.09} & 2.46 & 15.63 \\
12     & 34.29 & 25.92 & 17.66 & 8.59 & 5.39 & \textbf{2.31} & 15.69 \\
\midrule
\rowcolor{gray!20}
\multicolumn{8}{l}{\textbf{Dual}} \\
2,6    & 35.11 & 25.93 & 17.76 & 7.48 & 5.62 & 2.70 & 15.77 \\
\rowcolor{pastelviolet}4,8    & \textbf{33.14} & \textbf{24.61} & \textbf{16.24} & {7.42} & 5.38 & {2.38} & \textbf{14.86} \\
8,12   & 33.45 & 25.71 & \underline{16.26} & 7.73 & 5.26 & 2.54 & \underline{15.16} \\
\midrule
\rowcolor{gray!20}
\multicolumn{8}{l}{\textbf{Triple}} \\
-1,4,8 & {34.56} & 25.66 & 16.38 & 8.03 & 6.68 & 2.81 & 15.68 \\
4,8,12 & {33.27} & 26.61 & 17.55 & 7.85 & 6.31 & 2.90 & 15.75 \\
\bottomrule
\end{tabular}
}
\vspace{-0.5cm}
\end{table}


\noindent\textbf{Effect of VIB Layer Placement:} The placement of VIB modules within the LLM determines at which depth noise-corrupted representations are compressed. Inserting bottlenecks too early may interfere with low-level features, while inserting them too late leaves noise unmitigated. We evaluate single, dual, and triple bottleneck configurations at various layer positions, where each index refers to the transformer layer after which the VIB is applied. For these experiments, we reports results in noisy training paradigm and set $\beta=\frac{0.1}{H}$.  Layer index $-1$ denotes placement immediately after the audio encoder, where the projection layer serves as the VIB module with $\beta = 10^{-7}$ following~\cite{bai2025mitigating}. Moreover, due to shape mismatch between the output of the audio encoder and LLM hidden dimension size, we don't do the interpolation in this case. Results under babble noise are reported in Table~\ref{tab:vib_layers}. 

Single-layer configurations perform comparably across positions, with no placement offering a consistent advantage over the others. This suggests that a single bottleneck lacks sufficient capacity to suppress noise reliably, regardless of where it is placed. Notably, the $-1$ configuration, which places the bottleneck at the audio encoder output, achieves the best average WER among all single-layer configurations at $15.03\%$, suggesting that early compression at the encoder-LLM interface is more effective than any single LLM-layer insertion. However, it still falls short of the best dual configuration. Among dual configurations, layers (4, 8) yields the best overall performance, achieving the lowest average WER and the strongest gains under extreme noise, while remaining competitive at higher SNR levels. The (8, 12) configuration is a close second, but the advantage of (4, 8) at low SNR suggests that compressing representations at an earlier intermediate layer is beneficial when noise is most severe. Adding a third bottleneck degrades performance across all conditions, including clean speech. This indicates that stacking too many compression stages leads to over-regularisation.

\noindent\textbf{Effect of $\beta$:} The $\beta$ hyperparameter controls the strength of the KL regularisation term, trading off compression of noisy representations against retention of task-relevant content. Too small values reduces the regularisation effect, while too large values risk discarding task-relevant features. We ablate over four values normalised by the LLM hidden dimension $H$: $\{0.05, 0.1, 0.2, 1\}$, using the (4, 8) dual configuration identified above. Results are reported in Table~\ref{tab:beta_ablation}. At $\beta = 1$, performance degrades sharply across all SNR levels with average WER rising to $17.68\%$, confirming that excessive compression is harmful. Among the remaining values, $\beta = 0.1$ achieves the best average WER of $14.86\%$ and is the most consistent across SNR levels, and is therefore used in all other experiments.

\begin{table}[t]
\centering
\caption{WER (\%) under babble for $\beta$ values, which are normalized by $H$, hidden dimension of LLM backbone. The best results are in bold and the second-best are underlined.}
\label{tab:beta_ablation}
\scriptsize
\setlength{\tabcolsep}{4pt}
\renewcommand{\arraystretch}{1.2}

\resizebox{0.83\columnwidth}{!}{
\begin{tabular}{c *{7}{c}}
\toprule
& \multicolumn{6}{c}{SNR (dB)} & \\
\cmidrule(lr){2-7}
$\beta$ & $-10$ & $-5$ & $-2$ & $0$ & $5$ & $\infty$ & Avg\\
\midrule
$0.05$ & \underline{33.84} & 25.42 & \underline{16.52} & \underline{7.63} & \underline{5.48} & 2.84 & 15.29 \\
\rowcolor{pastelviolet}$0.1$  & \textbf{33.14} & \textbf{24.61} & \textbf{16.24} & \textbf{7.42} & \textbf{5.38} & \textbf{2.38} & \textbf{14.86} \\
$0.2$  & 34.40 & \underline{24.80} & 16.66 & 7.63 & 5.53 & \underline{2.49 }& \underline{15.25} \\
$1$    & 35.86 & 27.73 & 19.49 & 10.72 & 8.03 & 4.25 & 17.68 \\
\bottomrule
\end{tabular}
}
\end{table}


\begin{table}[t]
\centering
\caption{WER (\%) under babble for interpolation $\alpha$ values. The best results are in bold and the second-
best are underlined.}
\label{tab:beta_audio_encoder}
\scriptsize
\setlength{\tabcolsep}{4pt}
\renewcommand{\arraystretch}{1.2}

\resizebox{0.83\columnwidth}{!}{
\begin{tabular}{c *{7}{c}}
\toprule
& \multicolumn{6}{c}{SNR (dB)} & \\
\cmidrule(lr){2-7}
$\alpha$ & $-10$ & $-5$ & $-2$ & $0$ & $5$ & $\infty$ & Avg\\
\midrule
$0$ & 37.62 & 31.13 & 22.41 & 13.23 & 8.32 & 4.58 & 19.54 \\
$0\rightarrow0.5$ & \underline{33.54} & \underline{26.75} & \underline{18.45} & \underline{9.05} & \underline{6.56} & \underline{3.43} & \underline{16.29} \\
\rowcolor{pastelviolet}$0.5$ & \textbf{33.14} & \textbf{24.61} & \textbf{16.24} & \textbf{7.42} & \textbf{5.38} & \textbf{2.38} & \textbf{14.86} \\
\bottomrule
\end{tabular}
}
\vspace{-0.5cm}
\end{table}

\noindent\textbf{Role of Interpolation Coefficient $\alpha$.} We compare three configurations using the (4, 8) dual setup with $\beta = \frac{0.1}{H}$: setting $\alpha = 0$, cosine scheduling $\alpha$ from $0$ to $0.5$ following~\cite{ohvisual}, and fixing $\alpha = 0.5$ throughout training. Setting $\alpha = 0$ replaces the hidden state entirely with the sampled representation and leads to the worst performance across conditions, consistent with over-regularisation. The scheduled variant improves over $\alpha = 0$ but still shows higher WER across SNR levels relative to the fixed setting, which we attribute to the period of near-complete compression early in training that the LLM must recover from. 

\section{Conclusion}
We introduce VIB-AVSR, a method for improving noise robustness in audio-visual speech recognition by inserting Variational Information Bottleneck modules into intermediate layers of the LLM backbone. The VIB bottleneck applies compression to audio representations during training, encouraging the model to retain acoustically relevant content while discarding noise-correlated features. Experiments under babble and speech noise show consistent WER reductions over the Llama-AVSR baseline across SNR levels and training paradigms. Notably, the gains hold even when no noise augmentation is used during training, suggesting that the compression term  promotes representations that generalise better to noisy conditions.

\section{Generative AI Use Disclosure}
Generative AI tools were used solely for language editing and manuscript polishing. All research ideas, methodology, experiments, and conclusions were developed and are the responsibility of the authors.

\section{Acknowledgements}
All data processing and experiments were conducted at Imperial College London.

\bibliographystyle{IEEEtran}
\bibliography{mybib}

@inproceedings{bai2025mitigating,
  title={Mitigating hallucinations in large vision-language models by adaptively constraining information flow},
  author={Bai, Jiaqi and Guo, Hongcheng and Peng, Zhongyuan and Yang, Jian and Li, Zhoujun and Li, Mohan and Tian, Zhihong},
  booktitle={Proceedings of the AAAI Conference on Artificial Intelligence},
  volume={39},
  number={22},
  pages={23442--23450},
  year={2025}
}

@article{musan,
  title={Musan: A music, speech, and noise corpus},
  author={Snyder, David and Chen, Guoguo and Povey, Daniel},
  journal={arXiv preprint arXiv:1510.08484},
  year={2015}
}

@article{vaswani2017attention,
  title={Attention is all you need},
  author={Vaswani, Ashish and Shazeer, Noam and Parmar, Niki and Uszkoreit, Jakob and Jones, Llion and Gomez, Aidan N and Kaiser, {\L}ukasz and Polosukhin, Illia},
  journal={Advances in neural information processing systems},
  volume={30},
  year={2017}
}

@article{hu2022lora,
  title={Lora: Low-rank adaptation of large language models.},
  author={Hu, Edward J and Shen, Yelong and Wallis, Phillip and Allen-Zhu, Zeyuan and Li, Yuanzhi and Wang, Shean and Wang, Liang and Chen, Weizhu and others},
  journal={Iclr},
  volume={1},
  number={2},
  pages={3},
  year={2022}
}

@inproceedings{grattafiori2024llama,
  title={The Llama 3 herd of models},
  author={Grattafiori, Aaron and Dubey, Abhimanyu and Jauhri, Abhinav and Pandey, Abhinav and Kadian, Abhishek and Al-Dahle, Ahmad and Letman, Aiesha and Mathur, Akhil and Schelten, Alan and Vaughan, Alex and others},
  booktitle={Neural Information Processing Systems},
  year={2024},
  organization={Curran Associates}
}

@inproceedings{radford2023robust,
  title={Robust speech recognition via large-scale weak supervision},
  author={Radford, Alec and Kim, Jong Wook and Xu, Tao and Brockman, Greg and McLeavey, Christine and Sutskever, Ilya},
  booktitle={International conference on machine learning},
  pages={28492--28518},
  year={2023},
  organization={PMLR}
}

@inproceedings{son2017lip,
  title={Lip reading sentences in the wild},
  author={Son Chung, Joon and Senior, Andrew and Vinyals, Oriol and Zisserman, Andrew},
  booktitle={Proceedings of the IEEE conference on computer vision and pattern recognition},
  pages={6447--6456},
  year={2017}
}

@inproceedings{kingma2022autoencodingvariationalbayes,
      title={Auto-Encoding Variational Bayes}, 
      author={Diederik P Kingma and Max Welling},
      booktitle={International Conference on
Learning Representations (ICLR)},
      year={2014},
}

@inproceedings{ohvisual,
  title={Visual Instruction Bottleneck Tuning},
  author={Oh, Changdae and Li, Jiatong and Im, Shawn and Li, Sharon},
  booktitle={The Thirty-ninth Annual Conference on Neural Information Processing Systems},
  year={2025}
}

@article{tishby2000information,
  title={The information bottleneck method},
  author={Tishby, Naftali and Pereira, Fernando C and Bialek, William},
  journal={arXiv preprint physics/0004057},
  year={2000}
}

@inproceedings{cappellazzo2025scaling,
  title={Scaling and Enhancing LLM-based AVSR: A Sparse Mixture of Projectors Approach},
  author={Cappellazzo, Umberto and Kim, Minsu and Petridis, Stavros and Falavigna, Daniele and Brutti, Alessio},
  booktitle={Proc. Interspeech 2025},
  pages={1823--1827},
  year={2025}
}

@inproceedings{alemi2017deep,
  title={Deep Variational Information Bottleneck},
  author={Alemi, Alexander A and Fischer, Ian and Dillon, Joshua V and Murphy, Kevin},
  booktitle={International Conference on Learning Representations},
  year={2017}
}

@inproceedings{yeo2025mms,
  title={Mms-llama: Efficient llm-based audio-visual speech recognition with minimal multimodal speech tokens},
  author={Yeo, Jeong Hun and Rha, Hyeongseop and Park, Se Jin and Ro, Yong Man},
  booktitle={Findings of the Association for Computational Linguistics: ACL 2025},
  pages={20724--20735},
  year={2025}
}

@inproceedings{yeo2024visual,
  title={Where visual speech meets language: VSP-LLM framework for efficient and context-aware visual speech processing},
  author={Yeo, Jeonghun and Han, Seunghee and Kim, Minsu and Ro, Yong Man},
  booktitle={Findings of the Association for Computational Linguistics: EMNLP 2024},
  pages={11391--11406},
  year={2024}
}

@inproceedings{cappellazzo2025large,
  title={Large language models are strong audio-visual speech recognition learners},
  author={Cappellazzo, Umberto and Kim, Minsu and Chen, Honglie and Ma, Pingchuan and Petridis, Stavros and Falavigna, Daniele and Brutti, Alessio and Pantic, Maja},
  booktitle={ICASSP 2025-2025 IEEE International Conference on Acoustics, Speech and Signal Processing (ICASSP)},
  pages={1--5},
  year={2025},
  organization={IEEE}
}

@inproceedings{shi2022robust,
  title={Robust Self-Supervised Audio-Visual Speech Recognition},
  author={Shi, Bowen and Hsu, Wei-Ning and Mohamed, Abdelrahman},
  booktitle={Proc. Interspeech 2022},
  pages={2118--2122},
  year={2022}
}

@inproceedings{ma2023auto,
  title={Auto-avsr: Audio-visual speech recognition with automatic labels},
  author={Ma, Pingchuan and Haliassos, Alexandros and Fernandez-Lopez, Adriana and Chen, Honglie and Petridis, Stavros and Pantic, Maja},
  booktitle={ICASSP 2023-2023 IEEE International Conference on Acoustics, Speech and Signal Processing (ICASSP)},
  pages={1--5},
  year={2023},
  organization={IEEE}
}

@inproceedings{shilearning,
  title={Learning Audio-Visual Speech Representation by Masked Multimodal Cluster Prediction},
  author={Shi, Bowen and Hsu, Wei-Ning and Lakhotia, Kushal and Mohamed, Abdelrahman},
  booktitle={International Conference on Learning Representations}
}

@inproceedings{ma2021end,
  title={End-to-end audio-visual speech recognition with conformers},
  author={Ma, Pingchuan and Petridis, Stavros and Pantic, Maja},
  booktitle={ICASSP 2021-2021 IEEE International Conference on Acoustics, Speech and Signal Processing (ICASSP)},
  pages={7613--7617},
  year={2021},
  organization={IEEE}
}

@inproceedings{petridis2018audio,
  title={Audio-visual speech recognition with a hybrid ctc/attention architecture},
  author={Petridis, Stavros and Stafylakis, Themos and Ma, Pingchuan and Tzimiropoulos, Georgios and Pantic, Maja},
  booktitle={2018 IEEE Spoken Language Technology Workshop (SLT)},
  pages={513--520},
  year={2018},
  organization={IEEE}
}

@inproceedings{burchi2023audio,
  title={Audio-visual efficient conformer for robust speech recognition},
  author={Burchi, Maxime and Timofte, Radu},
  booktitle={Proceedings of the IEEE/CVF winter conference on applications of computer vision},
  pages={2258--2267},
  year={2023}
}

@inproceedings{rouditchenko2024whisper,
  title={Whisper-flamingo: Integrating visual features into whisper for audio-visual speech recognition and translation},
  author={Rouditchenko, Andrew and Gong, Yuan and Thomas, Samuel and Karlinsky, Leonid and Kuehne, Hilde and Feris, Rogerio and Glass, James},
  booktitle={Interspeech},
  year={2024}
}

@article{li2026noise,
  title={Noise-Robust AV-ASR Using Visual Features Both in the Whisper Encoder and Decoder},
  author={Li, Zhengyang and Graave, Thomas and M{\"o}ller, Bj{\"o}rn and Wu, Zehang and Franz, Matthias and Fingscheidt, Tim},
  journal={arXiv preprint arXiv:2601.18396},
  year={2026}
}

@article{cappellazzo2025mome,
  title={MoME: Mixture of Matryoshka Experts for Audio-Visual Speech Recognition},
  author={Cappellazzo, Umberto and Kim, Minsu and Ma, Pingchuan and Chen, Honglie and Liu, Xubo and Petridis, Stavros and Pantic, Maja},
  journal={arXiv preprint arXiv:2510.04136},
  year={2025}
}

@article{cappellazzo2025omni,
  title={Omni-AVSR: Towards Unified Multimodal Speech Recognition with Large Language Models},
  author={Cappellazzo, Umberto and Liu, Xubo and Ma, Pingchuan and Petridis, Stavros and Pantic, Maja},
  journal={arXiv preprint arXiv:2511.07253},
  year={2025}
}

@article{kim2025mohave,
  title={MoHAVE: mixture of hierarchical audio-visual experts for robust speech recognition},
  author={Kim, Sungnyun and Jang, Kangwook and Bae, Sangmin and Cho, Sungwoo and Yun, Se-Young},
  journal={arXiv preprint arXiv:2502.10447},
  year={2025}
}

@inproceedings{hong2023watch,
  title={Watch or listen: Robust audio-visual speech recognition with visual corruption modeling and reliability scoring},
  author={Hong, Joanna and Kim, Minsu and Choi, Jeongsoo and Ro, Yong Man},
  booktitle={Proceedings of the IEEE/CVF Conference on Computer Vision and Pattern Recognition},
  pages={18783--18794},
  year={2023}
}

@inproceedings{anand2026mitigating,
  title={Mitigating attention sinks and massive activations in audio-visual speech recognition with llms},
  author={Anand, Anand and Cappellazzo, Umberto and Petridis, Stavros and Pantic, Maja},
  booktitle={ICASSP 2026-2026 IEEE International Conference on Acoustics, Speech and Signal Processing (ICASSP)},
  pages={17942--17946},
  year={2026},
  organization={IEEE}
}

@inproceedings{cappellazzo2025adaptive,
  title={Adaptive audio-visual speech recognition via matryoshka-based multimodal llms},
  author={Cappellazzo, Umberto and Kim, Minsu and Petridis, Stavros},
  booktitle={2025 IEEE Automatic Speech Recognition and Understanding Workshop (ASRU)},
  pages={1--8},
  year={2025},
  organization={IEEE}
}

@article{rouditchenko2025mwhisper,
  title={mWhisper-Flamingo for Multilingual Audio-Visual Noise-Robust Speech Recognition},
  author={Rouditchenko, Andrew and Thomas, Samuel and Kuehne, Hilde and Feris, Rogerio and Glass, James},
  journal={IEEE Signal Processing Letters},
  year={2025},
  publisher={IEEE}
}

@inproceedings{kim2025multi,
  title={Multi-task corrupted prediction for learning robust audio-visual speech representation},
  author={Kim, Sungnyun and Cho, Sungwoo and Bae, Sangmin and Jang, Kangwook and Yun, Se-Young},
  booktitle={ICLR},
  year={2025}
}

@article{cappellazzo2026dr,
  title={Dr. SHAP-AV: Decoding Relative Modality Contributions via Shapley Attribution in Audio-Visual Speech Recognition},
  author={Cappellazzo, Umberto and Petridis, Stavros and Pantic, Maja},
  journal={arXiv preprint arXiv:2603.12046},
  year={2026}
}

@inproceedings{serdyuk2022transformer,
  title={Transformer-based video front-ends for audio-visual speech recognition for single and multi-person video},
  author={Serdyuk, Dmitriy and Braga, Otavio and Siohan, Olivier},
  booktitle={Interspeech},
  year={2022}
}

\end{document}